\documentclass[%
 reprint,
 showkeys,
superscriptaddress,
 amsmath,amssymb,
]{revtex4-1}
\usepackage{graphicx}
\usepackage{bm}
\usepackage{color,soul}
\usepackage{siunitx}\usepackage{bm}
\usepackage{siunitx}
\newcommand{\beginsupplement}{%
        \setcounter{table}{0}
        \renewcommand{\thetable}{S\arabic{table}}%
        \setcounter{figure}{0}
        \renewcommand{\thefigure}{S\arabic{figure}}%
     }

\usepackage{caption}
\usepackage{refstyle}

\begin{document}

\title{\bf Magneto-ionic control of spin polarization in magnetic tunnel junctions}

\author{Yingfen Wei}
\affiliation{Zernike Institute for Advanced Materials, University of Groningen, 9747 AG Groningen, The Netherlands }
\author{Sylvia Matzen}
\email{sylvia.matzen@u-psud.fr}
\affiliation{Centre for Nanoscience and Nanotechnology, CNRS UMR 9001, Universit\'e Paris-Sud, Universit\'e Paris-Saclay, 91120 Palaiseau, France}
\author{Cynthia P. Quinteros}
\affiliation{Zernike Institute for Advanced Materials, University of Groningen, 9747 AG Groningen, The Netherlands }
\author{Thomas Maroutian}
\affiliation{Centre for Nanoscience and Nanotechnology, CNRS UMR 9001, Universit\'e Paris-Sud, Universit\'e Paris-Saclay, 91120 Palaiseau, France}
\author{Guillaume Agnus}
\affiliation{Centre for Nanoscience and Nanotechnology, CNRS UMR 9001, Universit\'e Paris-Sud, Universit\'e Paris-Saclay, 91120 Palaiseau, France}
\author{Philippe Lecoeur}
\affiliation{Centre for Nanoscience and Nanotechnology, CNRS UMR 9001, Universit\'e Paris-Sud, Universit\'e Paris-Saclay, 91120 Palaiseau, France}
\author{Beatriz Noheda}
\email{b.noheda@rug.nl}
\affiliation{Zernike Institute for Advanced Materials, University of Groningen, 9747 AG Groningen, The Netherlands }
\affiliation{CogniGron center, University of Groningen, 9747 AG Groningen, The Netherlands }

\begin{abstract}
Magnetic tunnel junctions (MTJs) with Hf$_{0.5}$Zr$_{0.5}$O$_2$ barriers are reported to show both tunneling magnetoresistance effect (TMR) and tunneling electroresistance effect (TER), displaying four resistance states by magnetic and electric field switching. Here we show that, under electric field cycling of large enough magnitude, the TER can reach values as large as 10$^6$\%. Moreover, concomitant with this TER enhancement, the devices develop electrical control of spin polarization, with sign reversal of the TMR effect. Currently, this intermediate state exists for a limited number of cycles and  understanding the origin of these phenomena is key to improve its stability. The experiments presented here point to the magneto-ionic effect as the origin of the large TER and strong magneto-electric coupling, showing that ferroelectric polarization switching of the tunnel barrier is not the main contribution.
\end{abstract}
\today
\maketitle

\label{section:Intro}
Combining the TMR effect of magnetic tunnel junctions (MTJs) with additional functionalities provided by the tunnel barrier, i.e. using multiferroic\cite{gajek2007tunnel} or ferroelectric\cite{rodriguez2003resistive, tsymbal2012ferroelectric} layers as barriers, has drawn considerable attention driven by their potential application in multilevel memories. In these devices, four resistance states are achieved by means of both the TMR (resistance change induced by magnetic field switching) and the TER (resistance change by electric switching) effects.\cite{scott2007data, hur2004electric,
bibes2008multiferroics} In addition, by combining two ferroic orders (ferromagnetic and ferroelectric), the coupling between the magnetic and electric degrees of freedom could realize electric field controlled spintronics, promising for the development of low-power and fast devices.\cite{ortega2015multifunctional, webster2002wiley,eerenstein2006multiferroic, ramesh2007multiferroics, fiebig2005revival, 5garcia2010ferroelectric, yin2017review, bibes2008multiferroics}

Four main types of magnetoelectric (ME) coupling mechanisms\cite{2dong2019magnetoelectricity, dong2015multiferroic, bauer2015magneto} are established. Firstly, spin-orbit coupling, which can directly link the breaking of space and time inversion symmetry (charge dipoles and magnetic moments).\cite{eerenstein2006multiferroic} Secondly, spin-lattice coupling,\cite{lee2010strong, fabreges2009spin} that profits from the piezoelectric properties of ferroelectrics and the magnetostrictive properties of ferromagnets. In this case strain can couple electric field with magnetization or magnetic field with electrical polarization.\cite{tsymbal2006tunneling, gajek2007tunnel} Thirdly, the ME coupling can originate from spin-charge coupling mediated by the carrier density.\cite{3rondinelli2008carrier} At the interface between an insulator and a ferromagnetic metal, accumulation of spin-polarized carriers and, thus magnetization, is expected upon application of an electric field that leads to polarization of the dielectric. This effect is enhanced in the case of a polar barrier, as a larger number of carriers is necessary for screening. Finally, the magneto-ionic effect\cite{bauer2015magneto} has been recently proposed, by which the applied electric field induces ion migration that modifies the interfaces of the ferromagnetic layers involved in the junctions. All these mechanisms could contribute to the ME coupling, separately or jointly.

In this work, tunnel barriers of crystalline Hf$_{0.5}$Zr$_{0.5}$O$_2$ (HZO) are used in MTJs. Crystalline HZO grown under certain conditions has shown nanoscale ferroelectricity.\cite{boscke2011ferroelectricity, park2015ferroelectricity}. Epitaxial growth of crystalline HZO can also be achieved\cite{shimizu2016demonstration} and has been recently also demonstrated on perovskite substrates with La$_{0.7}$Sr$_{0.3}$MnO$_3$ (LSMO) as bottom elecrode\cite{4wei2018rhombohedral,lyu2018robust, lyu2019growth}. Four resistance states have been obtained in this type of junctions by both magnetic and electric field switching, but no ME coupling was reported.\cite{10wei2019magnetic} Here we show that electric field cycling induces irreversible changes in the junction, which evolves from a negligible ME coupling state into a large ME coupling state. In the latter, sign reversal of the TMR effect is achieved by electrical switching reversibly. Concomitantly, with increasing number of cycles, the TER increases to values up to 10$^6$\%. In the following we discuss the mechanisms that lead to such phenomena.

\begin{figure*}
	\includegraphics[width=16cm]{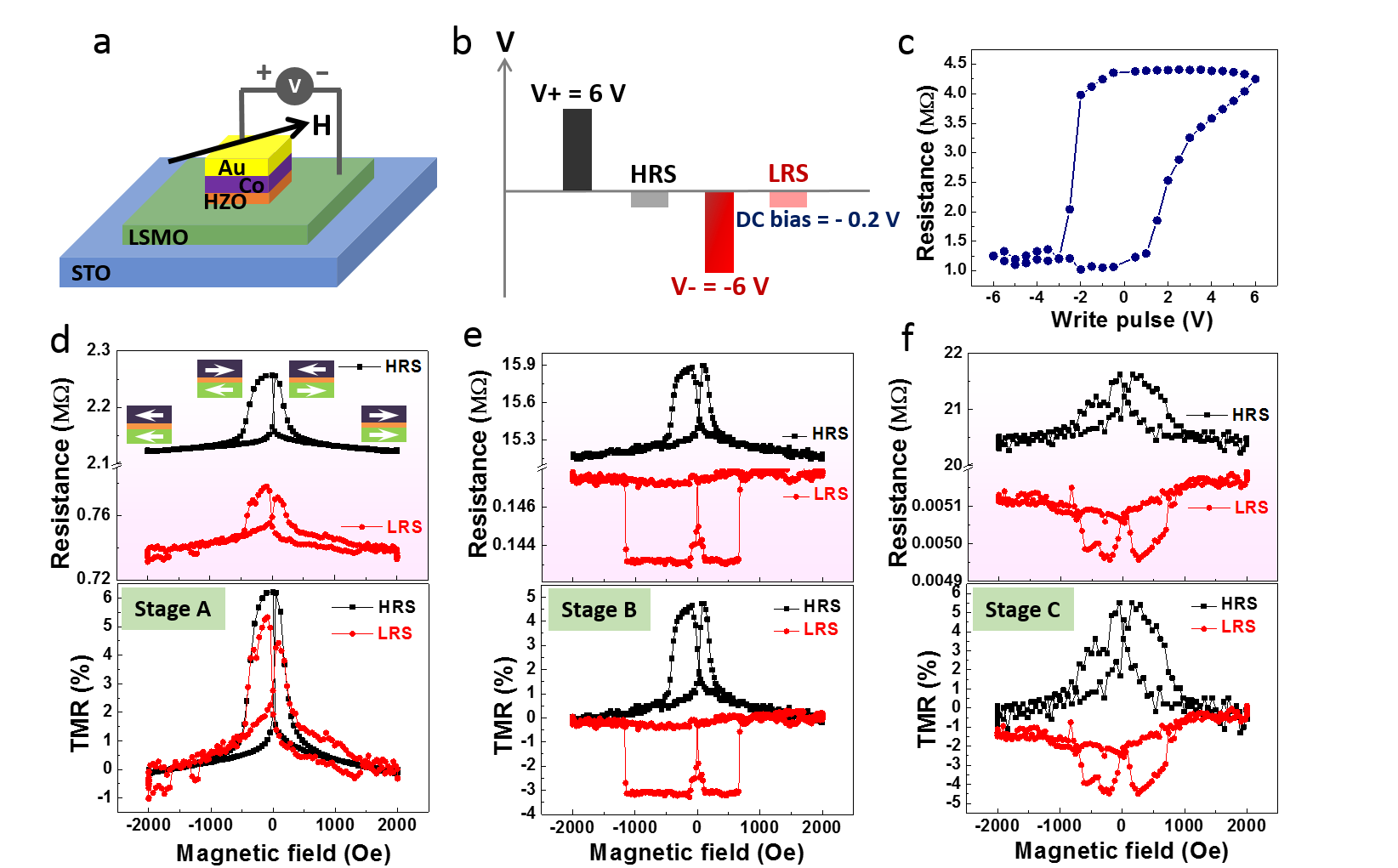}
	\caption{\label{fig:TMRdiffstage} 
a, Schematic drawing of a tunnel junction device, with bottom electrode grounded. The heterostructure is grown on a 001-oriented SrTiO$_3$ substrate. The applied magnetic field is along the [110] direction; b, Electrical pulses of 6 V with both positive (black) and negative (red) polarity and 500 $\mu s$ duration are applied to the junctions in order to bring them into the high (HRS) and low resistance state (LRS), respectively. All TMR loops are measured under a DC bias of -0.2 V, both in the HRS and LRS; c, Changes in resistance under the application of different amplitude of electrical pulses in the same junction, as shown in Fig. 3a of ref.\cite{10wei2019magnetic}; d, e, and f, resistance as a function of sweeping magnetic field (up panels) and TMR ratio (down panels) in the HRS (black squares) and LRS (red circles) measured at three different intermediate stages (named as stage A, B and C, respectively) upon repeated application of +/-6V electric pulses. Measurements shown here are performed at 50 K on a junction device with an electrode area of $30\mu m\times30\mu m$.}
\end{figure*}

MTJs have been fabricated by integrating 2 nm HZO tunnel barriers between top Co and bottom La$_{0.7}$Sr$_{0.3}$MnO$_3$ (LSMO) ferromagnetic (FM) electrodes. The HZO is highly crystalline and epitaxially grown on the LSMO electrode, which is in turn epitaxially grown on 001-oriented SrTiO$_3$ (STO) substrates. As reported in ref.\cite{10wei2019magnetic}, the large band gap and high resistance of the HZO layer allows to fabricate full devices with extended electrodes for wire bonding, despite the low thickness of the barrier. This is not possible with perovskite ferroelectric (FE) tunnel barriers with such small thickness and, thus, so far these devices have been limited to investigation by scanning probes. The schematic drawing of the devices used in the present work is shown in Fig. \ref{fig:TMRdiffstage}a. (See details in methods section). 

\begin{figure*}
	\includegraphics[width=10cm]{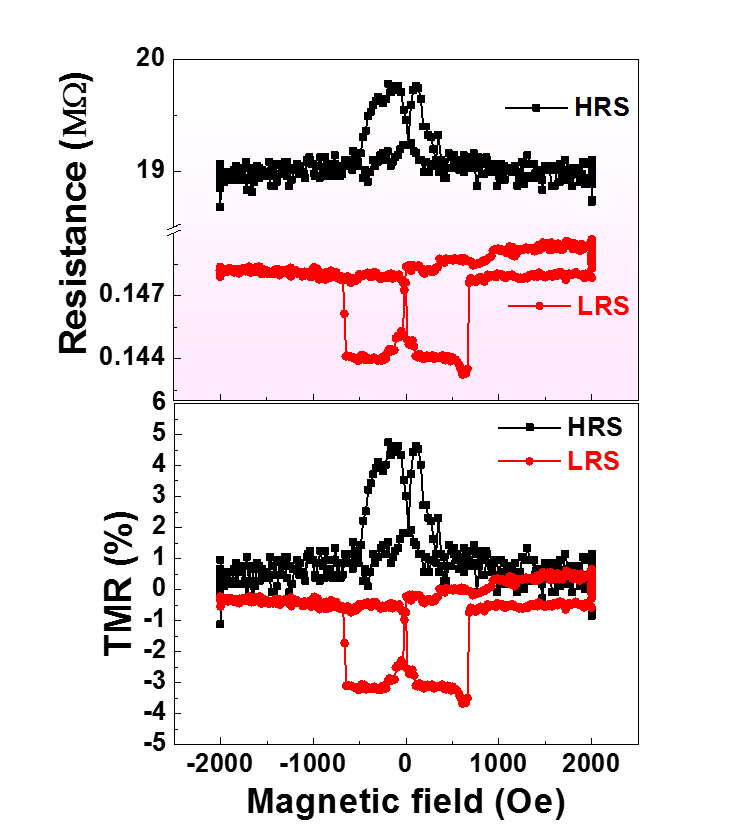}
	\centering 
	\caption{\label{fig:stablity} 
Second run of electrical switching of spin polarization at stage B. 
}
\end{figure*}

By the electrical pulse switching protocol shown in Fig.\ref{fig:TMRdiffstage}b, the junction switches between the high resistance state, HRS ($R_H$, after V+ pulse) and the low resistance state, LRS ($R_L$, after V- pulse). A voltage pulse with amplitude as large as 6 V is used in order to obtain the maximum resistance contrast (TER $\sim$ 400$\%$) (see Fig.\ref{fig:TMRdiffstage}c and ref.\cite{10wei2019magnetic}). In both HRS and LRS, TMR loops are obtained, as shown in Fig. \ref{fig:TMRdiffstage}d, leading to four resistance states ($R_{H_{\uparrow\uparrow}}$, $R_{H_{\uparrow\downarrow}}$, $R_{L_{\uparrow\uparrow}}$, $R_{L_{\uparrow\downarrow}}$, where the arrows signal the relative orientation of the electrodes magnetization). During the first few cycles, the TMR effect of the HRS ($\sim6.2\%$) and LRS ($\sim5.4\%$) are similar in magnitude (see Fig. \ref{fig:TMRdiffstage}d), indicating a negligible ME coupling, which differs from the strong coupling reported in perovskite tunnel barriers \cite{5garcia2010ferroelectric, 6pantel2012reversible, yin2013enhanced}. This stage, which we name stage A, is the one reported in ref.\cite{10wei2019magnetic}. 
Interestingly, after a few tens of cycles, the behaviour changes substantially, reaching the stage B, as shown in Fig. \ref{fig:TMRdiffstage}e: the TMR sign is reversed from positive (HRS) to negative (LRS) indicating that the spin polarization is switched by the external electric field in a reversible manner, as shown in Fig. \ref{fig:stablity} (see also supplementary information). In addition, the coercive field of the harder ferromagnet (the Co layer) in the LRS (with negative TMR) increases by, approximately, a factor of two, compared to the switching fields of the HRS (with positive TMR). Moreover, the switching becomes sharper in the LRS. The increase of the coercive field and steep switching of the Co layer upon electrical cycling could originate in a modification of the HZO/Co interface.\cite{bauer2015magneto, bauer2013voltage}. The number of cycles needed to reach the stage B has been found to differ depending on the junction under investigation. 

With further electric cycling (stage C), the TMR signal becomes more noisy, as observed in Fig. \ref{fig:TMRdiffstage}f. The switching magnetic fields for the direct and reversed TMR become comparable but still higher than those of stage A (Fig. \ref{fig:TMRdiffstage}d). However, the magnitude of the TMR effect is not substantially altered. In the meantime, the two magnetic states are less well defined with less abrupt magnetic switching than the previous two stages, which could indicate an increasing number of defects introduced in the stack. For longer cycling time, with number of cycles depending on the junction, the TMR effect eventually disappears but the TER effect is still present.

The TMR sign has been reported to reverse by modification of the Co interface, either by adding an interface layer\cite{de1999roleofinterface,leclair2001sign}, or by the electric field-controlled hybridization at the interface (modified by the different ferroelectric polarization states)\cite{6pantel2012reversible}. These mechanisms affecting the Co/barrier interface are consistent with the changes, described above, of the magnetization switching of the Co layer upon electric cycling. However, since the reversed TMR is not observed in the A-stage (it only appears upon repeated electric cycling), the ferroelectric polarization switching can be discarded as the main contribution to the TMR sign reversal. Moreover, changes in the junctions by the introduction of oxygen vacancies (V$_{O}^{2+}$) have also been reported to promote TMR sign reversal\cite{marun2007tunneling}. Given the possibility for positively charged V$_{O}^{2+}$ to migrate back and forth under the application of the electric field pulses with opposite polarities, we propose that ionic exchange is responsible for the observed changes of spin polarization, as well as the modification of the Co-HZO interface upon cycling.\cite{bauer2015magneto, bauer2012magnetoelectric, bauer2012electric}. 

\begin{figure}[tb]
	\includegraphics[width=10cm]{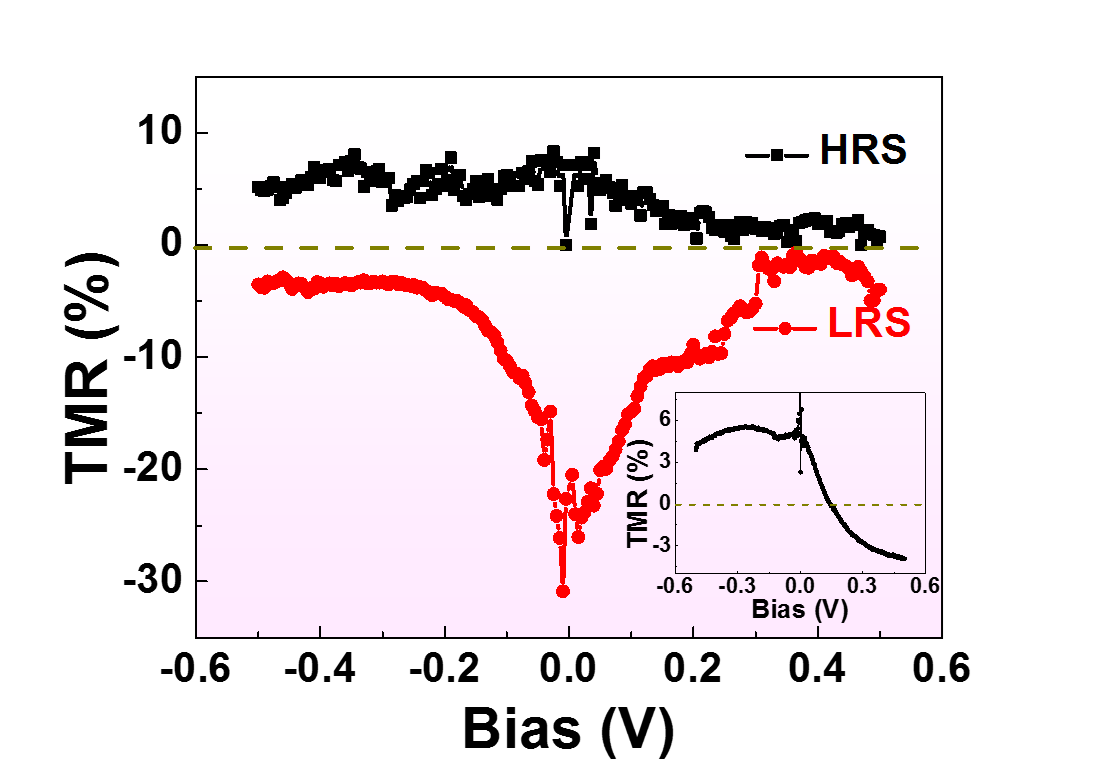}
	\caption{\label{fig:BiasreverseTMR} 
Bias dependence of TMR measured at stage C for the HRS and LRS, showing electrical switching of spin polarization, on the device shown in Fig. \ref{fig:TMRdiffstage}f, with size of $30\mu m\times30\mu m$, measured at 50 K. The inset shows the bias dependence of TMR in the as-grown state (before any electrical cycling) on the same device. Similar curves are obtained in both HRS and LRS at stage A for a $20\mu m\times20\mu m$ junction, as shown in ref. \cite{10wei2019magnetic}.}
\end{figure}

\begin{figure*}
	\includegraphics[width=15cm]{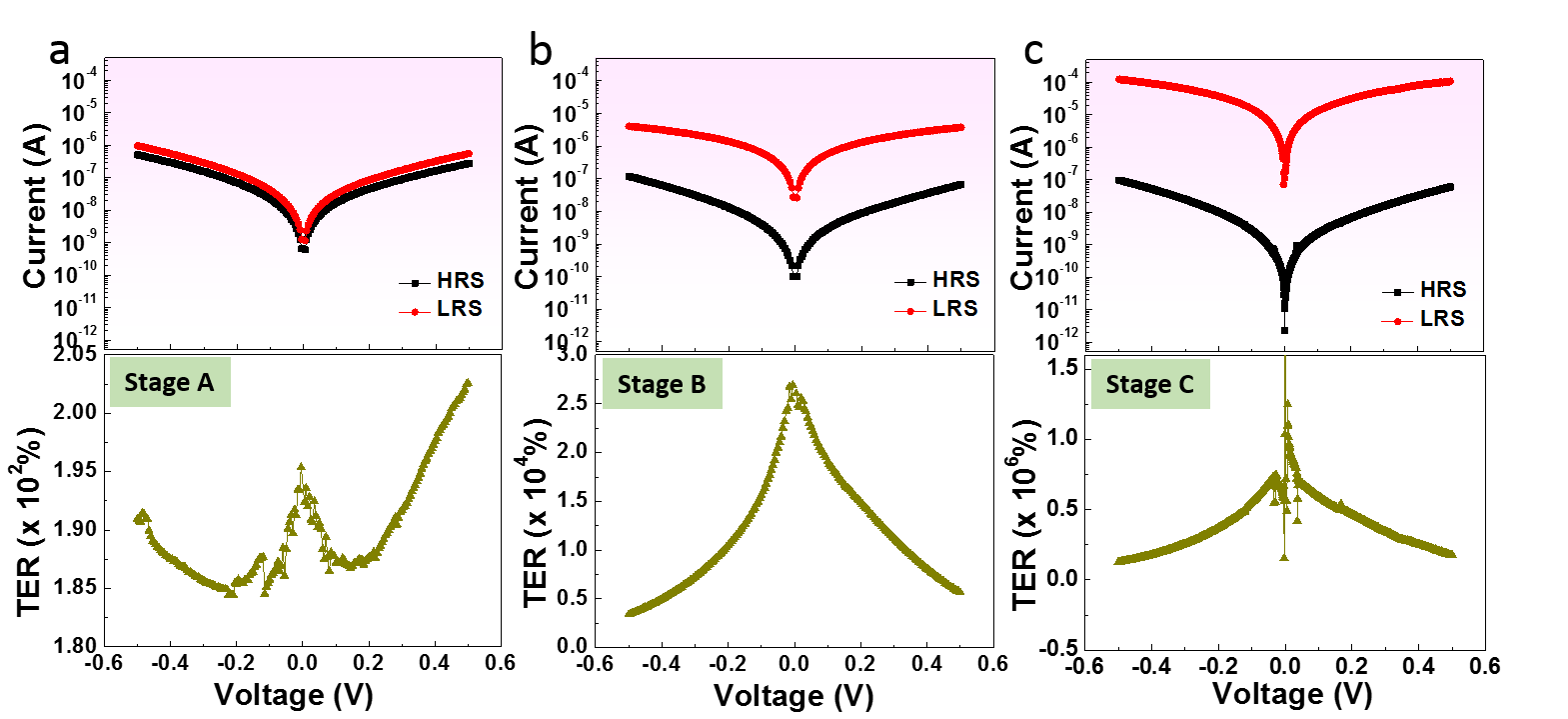}
	\centering 
	\caption{\label{fig:TERdiffstage} 
a, b and c, Current as a function of bias (up panel) in the HRS (black squares) and LRS (red circles) and TER values (down panel) at three different stages A, B, and C, which correspond to Fig.\ref{fig:TMRdiffstage}d-f. All are measured on a $30\mu m\times30\mu m$ junction at 50 K. 
	}
\end{figure*}

Focusing on stage C, from the I-V curves measured in parallel and anti-parallel magnetic states, we plot the bias dependence of 
the TMR for both HRS and LRS in Fig. \ref{fig:BiasreverseTMR}. A striking feature is that in HRS state the TMR exhibits a very weak bias-dependence and is always positive; while in the LRS, the TMR is always negative with a rapid drop of TMR with increasing bias (absolute value), characteristic of thin-film MTJs and attributed to spin-flip scattering\cite{moodera1999spin}. The electric field switching of spin polarization is, thus, evidenced over the whole investigated voltage range. Tuning of the read voltage allows to select the magnitude of the TMR change (e.g. Fig. \ref{fig:TMRdiffstage}f for -0.2 V read voltage). Looking at the bias-dependence of the TMR in the as-grown state for the same device (Fig. \ref{fig:BiasreverseTMR}, inset), and noting that similar curves are obtained in both HRS and LRS at stage A for different junctions \cite{10wei2019magnetic}, it is clear that electric field cycling completely changes the control of the spin polarization of the tunneling electrons. While the initial stage A shows a read voltage-controlled TMR sign change, already reported for Co-based junctions\cite{de1999inverseTMR, 38tsymbal2003resonant,10wei2019magnetic}, in stage C the TMR sign is wholly determined by the switchable resistance state of the device.

Concomitantly, the resistance ratio between the HRS and LRS (TER) also changes substantially during electric field cycling. By measuring the current-voltage (I-V) curves after positive and negative electric pulses, we can extract the TER at different bias by measuring the current ratio of HRS and LRS ($I_{L}/I_{H}$). TER rises from $10^2\%$ to $10^6\%$ (stage A to C) with a large number of intermediate states, as shown in Figs. \ref{fig:TERdiffstage}a-c, corresponding to Figs. \ref{fig:TMRdiffstage}d-f, respectively. 

Thus, it is shown that the junctions are strongly affected by the very large electric fields applied across the ultrathin HZO barrier, which induce stage B and C with highly enhanced magnetoelectric coupling and very large TER, of great interest for devices. The driving voltages required to achieve these stages are close to the junction breakdown field. Therefore, the ability to keep cycling the device with such a large stimulus could be due to a voltage drop somewhere in the device, such as at the Co-HZO interface. Understanding the mechanisms leading to this evolution would crucially help finding the optimal conditions required for applications (typically 10$^4$-10$^5$ cycles \cite{bez2003introduction} for flash memory, and much higher endurance in other non-volatile memories, such as ferroelectric memory, magnetoresistive memory, resistive memory, etc.\cite{boukhobza2017flash}).

\begin{figure*}
	\includegraphics[width=15cm]{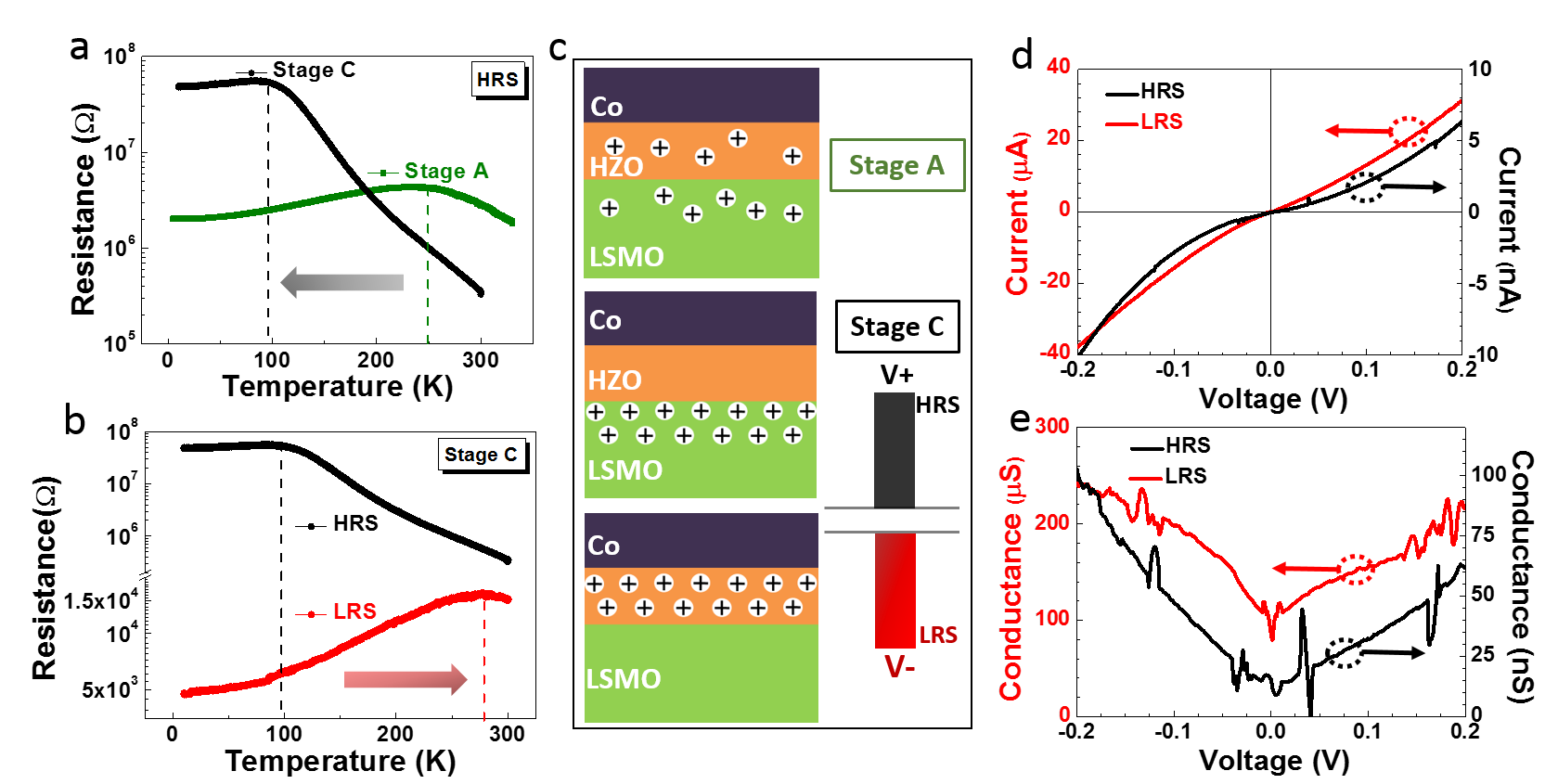}
	\centering 
	\caption{\label{fig:RTdiffstage} 
a. R-T curves in the HRS at stage A (green) and stage C (black), respectively; b, R-T curves in the HRS (black) and LRS (red) at the stage C; c, sketch of the proposed model of interface ionic exchange; d, e, I-V and dI/dV-V curves, respectively, in the HRS and LRS of stage C measured at 50 K. The spikes observed in the dI/dV curves are a consequence of the small experimental deviations in the experimental I-V data. Notice the different current/conductance scales in the LRS (left axis) and HRS (right axis).  
	}
\end{figure*}

To shed light into the factors affecting the evolution from stage A to C by electric cycling, transport measurements of resistance \textit{versus} temperature (R-T) are shown in Fig. \ref{fig:RTdiffstage}a. The same junction is measured in the HRS in stage A (green) and stage C (black). In stage A, a metal-insulator transition happens at around 250 K. This is the temperature at which the ferromagnetic/metal-to-paramagnetic/insulator transition of LSMO at the interface with HZO takes place and, thus, where the TMR disappears.\cite{10wei2019magnetic} Upon electric field cycling, the transition temperature decreases. In Fig. \ref{fig:RTdiffstage}a, the resistance of stage C (black) is shown to display the transition at around 100 K, which again coincides with the temperature at which TMR disappears (see Supplementary Fig. \ref{fig:TMR-Tdependence}). The decrease of transition temperature from stage A to C is consistent with an oxygen deficiency at  the LSMO interface\cite{cauro2001persistent, ge2015metal,marun2007tunneling} that increases with repeated electric field cycling. In addition, the junction $R_H$ increases from stage A to C (see Fig. \ref{fig:TMRdiffstage} and \ref{fig:TERdiffstage}), which also agrees with an increasing content of oxygen vacancies in the LSMO layer at the HRS upon cycling, since oxygen vacancies are well known to reduce the carrier (hole) concentration in LSMO.\cite{ge2015metal, schlueter2012evidence, yao2017direct, cauro2001persistent} 

Furthermore, the R-T measurements at stage C (with large TER and strong ME coupling) in the HRS and LRS are shown in Fig. \ref{fig:RTdiffstage}b. The transition temperature at the HRS (black), which had been lowered by the action of electric cycling to $\sim$ 100 K, increases up to $\sim$ 275 K, after the junction is brought to the LRS (red), which is higher than the transition temperature of the stage A ($\sim$ 250 K, see Fig. \ref{fig:RTdiffstage}a). This indicates that by applying a large negative pulse to the junction, the LSMO layer can reach an oxygen content larger than that of the initial stage. This is consistent with ionic exchange of oxygen vacancies in between the LSMO electrode and the HZO barrier during cycling, as represented in Fig. \ref{fig:RTdiffstage}c. Giant resistive switching by oxygen vacancies migration has also been observed in different ferroelectric oxides tunnel barriers\cite{qin2016resistive}.

In Fig. \ref{fig:RTdiffstage}c, we illustrate this possible scenario: in the as-grown state, both the LSMO and the HZO layers contain V$_{O}^{2+}$ (top panel). Upon electric field cycling, V$_{O}^{2+}$ are driven back and forth across the barrier. The evolution of the TER from $10^2$\% to $10^6$\% could be explained by the accumulation of the oxygen vacancies at the vicinity of the HZO/LSMO interface, thus increasing the V$_{O}^{2+}$ concentration that participates in the ionic exchange process. In this picture, the HRS is due to the oxygen vacancies being pushed into the LSMO electrode, resulting in a very resistive 
HZO/La$_{0.7}$Sr$_{0.3}$MnO$_{3-\delta}$ contact. The LRS is obtained with the oxygen vacancies drifting back into the HZO barrier upon negative voltage pulse application, greatly reducing the resistivity of the junction\cite{waser2009redox}. This gives rise to highly different current levels between HRS and LRS (large TER) as shown in Fig. \ref{fig:TERdiffstage}c and Fig. \ref{fig:RTdiffstage}d. Still, for both states, the non-linear I-V curves are similar (Fig. \ref{fig:RTdiffstage}d) and the shape of the differential conductance curves (Fig. \ref{fig:RTdiffstage}e) is compatible with tunneling conduction\cite{o2000colossal}, ruling out a drastic change of the conduction mechanism as probed by the 
investigated range of applied voltage.

An open question is the role of the ferroelectric polarization switching in these devices. Resistive switching by electric field has been reported in a wide variety of oxides\cite{sawa2008resistive, waser2010nanoionics, waser2009redox}, including binary oxides.\cite{seo2004reproducible, simmons1967new, choi2005resistive} In the case of ferroelectric tunnel barriers, the profile of the electronic barrier can be modified by polarization reversal, thus causing strong TER effect. \cite{rodriguez2003resistive, gruverman2009tunneling} However, in our case, polarization switching is not the main contribution to the large TER, since it increases upon cycling from stage A to C, as shown in Fig. \ref{fig:TERdiffstage}. It is interesting to notice that LSMO/HZO/Pt junctions, fabricated by I. Fina et al. \cite{Fina2019} with the same material as tunnel barrier but with a double barrier thickness, also show TER values of around 400$\%$. This TER is reproducible with cycling under relatively smaller driving voltages (4 V), suggesting that this could be the contribution of the ferroelectric polarization. With higher driving voltage, larger TER similar to those reported here, are observed.\cite{Fina2019}

In conclusion, TER values of up to $10^6\%$ coexisting with large ME coupling, by which the sign of the TMR effect is reversed with the electric field switching, have been achieved after cycling of Co/HZO/LSMO tunnel junctions with large enough electric fields. The temperature dependence of the transport behaviour is consistent with the exchange of oxygen vacancies at the LSMO/HZO interface, together with modifications of the HZO/Co interface. Next, an electrical protocol needs to be designed in order to increase the endurance of this state.\\     

\section*{Methods}
Thin films of Hf$_{0.5}$Zr$_{0.5}$O$_2$ (HZO) barrier with thickness of 2 nm were grown by pulsed laser deposition (PLD) on FM La$_{0.7}$Sr$_{0.3}$MnO$_3$(LSMO)-buffered (001)-SrTiO$_3$ substrates. The thickness of LSMO film is around 30 nm. Details of the growth conditions can be found in ref.\cite{4wei2018rhombohedral}. 50 nm FM Cobalt with a protective layer of Au (50nm), to preserve Co from oxidation, were deposited by sputtering on top of the HZO layer, to form the LSMO (FM) / HZO (FE)/ Co (FM) stacks. Junctions with different sizes, ranging from $10 \mu m \times 10 \mu m$ to $30 \mu m \times 30 \mu m$, are fabricated (see details in ref.\cite{10wei2019magnetic}). The electrical measurements are performed using a Keithley 237 source measurement unit and a Keithley 4200A-SCS parameter analyzer, and the temperature environment and magnetic field are supplied by a Physical Properties Measurement System (PPMS) by Quantum Design. As shown in the schematic drawing in Fig. \ref{fig:TMRdiffstage}a, the voltage source is applied on the LSMO/HZO/Co stack with bottom electrode grounded (for a positive bias, the electrons are tunneling from LSMO to Co). The magnetic field is swept along the easy magnetization axis of LSMO in the [110] direction.\\

\section*{Acknowledgements}
The authors are grateful to Tamalika Banerjee and Manuel Bibes for useful discussions. YW and BN acknowledge a China Scholarship Council grant and a Van Gogh travel grant.

\bibliographystyle{aipnum4-1} 
\bibliography{Ref}

\clearpage

\beginsupplement

\begin{figure*}
	\includegraphics[width=12cm]{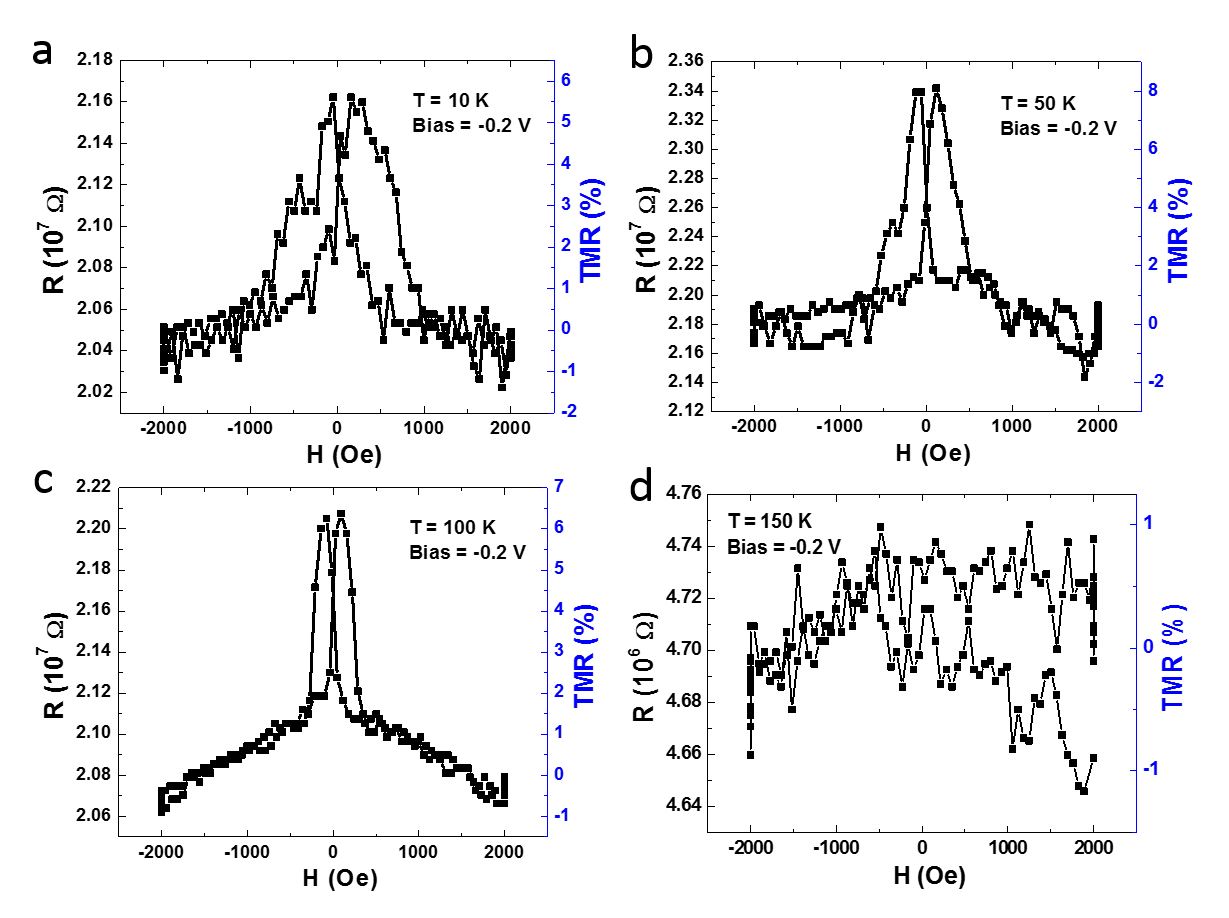}
	\centering 
	\caption{\label{fig:TMR-Tdependence} 
Temperature dependent TMR loops of HRS at stage C. a, 10 K; b, 50 K; c, 100 K; d, 150 K.
	}
\end{figure*}

\end{document}